\documentclass[aps,prb,twocolumn,showpacs]{revtex4}

\usepackage{graphicx}
\usepackage{amsmath}
\usepackage{amssymb}

\begin{document}

\title{Thermodynamic properties in triangular-lattice superconductors}

\author{Xixiao Ma and Ling Qin}

\affiliation{Department of Physics, Beijing Normal University, Beijing 100875, China}

\author{Huaisong Zhao}

\affiliation{College of Physics, Qingdao University, Qingdao 266071, China}

\author{Yu Lan}

\affiliation{College of Physics and Electronic Engineering, Hengyang Normal University, Hengyang 421002, China}

\author{Shiping Feng}

\affiliation{Department of Physics, Beijing Normal University, Beijing 100875, China~~}

\begin{abstract}
The study of superconductivity arising from doping a Mott insulator has become a central issue in the area of superconductivity. Within the framework of the kinetic-energy-driven superconducting mechanism, we discuss the thermodynamic properties in triangular-lattice superconductors. It is shown that a sharp peak in the specific-heat appears at the superconducting transition temperature $T_{\rm c}$, and then the specific-heat varies exponentially as a function of temperature for the temperatures $T<T_{\rm c}$ due to the absence of the d-wave gap nodes at the charge-carrier Fermi surface. In particular, the upper critical field follows qualitatively the Bardeen-Cooper-Schrieffer type temperature dependence, and has the same dome-shaped doping dependence as $T_{c}$.
\end{abstract}
\pacs{74.25.Bt, 74.20.Mn, 74.62.Dh, 74.70.-b}


\maketitle

\section{Introduction}

The parent compounds of cuprate superconductors are identified as Mott insulators \cite{Bednorz86,Kastner98}, in which the lack of conduction arises from the strong electron-electron repulsion. Superconductivity then is obtained by adding charge carriers to insulating parent compounds with the superconducting (SC) transition temperature $T_{\rm c}$ has a dome-shaped doping dependence \cite{Tallon95}. Since the discovery of superconductivity in cuprate superconductors, the search for other families of superconductors that might supplement what is known about the SC mechanism of doped Mott insulators has been of great interest. Fortunately, it has been found \cite{Takada03} that there is a class of cobaltate superconductors Na$_{x}$CoO$_{2}\cdot y$H$_{2}$O, which displays most of the structural and electronic features thought to be important for superconductivity in cuprate superconductors: strong two-dimensional character, proximity to a magnetically ordered nonmetallic state, and electron spin 1/2. In particular, $T_{\rm c}$ in cobaltate superconductors has the same unusual dome-shaped dependence on charge-carrier doping \cite{Schaak03,Milne04,Sakurai06,Michioka06}. However, there is one interesting difference: in the cuprate superconductors \cite{Bednorz86,Kastner98}, Cu ions in a square array are ordered antiferromagnetically, and then spin fluctuations are thought to play a crucial role in the charge-carrier pairing, while in the cobaltate superconductors \cite{Takada03,Schaak03,Milne04,Sakurai06,Michioka06}, Co ions in a triangular array are magnetically frustrated, and therefore this geometric frustration may suppress $T_{\rm c}$ to low temperatures. It has been argued that the triangular-lattice cobaltate superconductors are probably the only system other than the square-lattice cuprate superconductors where a doped Mott insulator becomes a superconductor.

The heat-capacity measurement of the specific-heat can probe the bulk properties of a superconductor, which has been proven as a powerful tool to investigate the low-energy quasiparticle excitations, and therefore gives information about the charge-carrier pairing symmetry, specifically, the existence of gap nodes at the Fermi Surface \cite{Schrieffer64}. In conventional superconductors \cite{Schrieffer64}, the absence of the low-energy quasiparticle excitations is reflected in the thermodynamic properties, where the specific-heat of conventional superconductors is experimentally found to be exponential at low temperatures, since conventional superconductors are fully gaped at the Fermi surface. However, the situation in the triangular-lattice cobaltate superconductors is rather complicated, since the experimental results obtained from different measurement techniques show a strong sample dependence \cite{Jin03,Ueland04,Lorenz04,Yang05,Jin05,Sasaki04,Maska04,Chou04,Oeschler08,Oeschler08b,Kobayashi03,Fujimoto04,Ihara06,Kato06,Zheng06}. Thus it is rather difficult to obtain conclusive results. The early specific-heat measurements \cite{Lorenz04,Yang05,Jin05} showed that the specific-heat in the triangular-lattice cobaltate superconductors reveals a sharp peak at $T_{\rm c}$, and can be explained phenomenologically within the Bardeen-Cooper-Schrieffer (BCS) formalism under an unconventional SC symmetry with line nodes. However, by contrast, the latest heat-capacity measurements \cite{Oeschler08,Oeschler08b} indicated that among a large number of the gap symmetries that have been suggested \cite{Mazin05,Lee90}, the SC-state with d-wave ($d_{1}+id_{2}$ pairing) symmetry without gap nodes at the Fermi surface is consistent with the observed specific-heat data. Furthermore, by virtue of the magnetization measurement technique, the value of the upper critical field and its temperature dependence have been observed for all the temperatures $T\leq T_{c}$ \cite{Jin05,Sasaki04,Maska04,Chou04}, where the temperature dependence of the upper critical field follows qualitatively the BCS type temperature dependence. On the theoretical hand, there is a general consensus that superconductivity in the triangular-lattice cobaltate superconductors is caused by the strong electron correlation \cite{Baskaran03,Kumar04,Wang04,Ogata03,Liu05,Qin15,Chen13,Valkov15}. Using the resonating-valence-bond mean-field approach, it has been suggested that the spin fluctuation enhanced by the dopant dynamics leads to a d-wave SC-state \cite{Baskaran03,Kumar04,Wang04}. Based on the mean-field variational approach with Gutzwiller approximation, a d-wave SC-state is realized in the parameter region close to the triangular-lattice cobaltate superconductors \cite{Ogata03}. Within the framework of the kinetic-energy-driven SC mechanism \cite{Feng0306,Feng15}, it has been demonstrated that charge carriers are held together in d-wave pairs at low temperatures by the attractive interaction that originates directly from the kinetic energy by the exchange of spin excitations \cite{Liu05,Qin15}. Moreover, superconductivity with the d-wave symmetry has been explored by a large-scale dynamical cluster quantum Monte Carlo simulation on the triangular-lattice Hubbard model \cite{Chen13}. In particular, using the diagram technique in the atomic representation, the SC phase with the d-wave symmetry in an ensemble of the Hubbard fermions on a triangular lattice has been discussed \cite{Valkov15}, where the domelike shape of the doping dependence of $T_{\rm c}$ is obtained. However, to the best of our knowledge, the thermodynamic properties of the triangular-lattice cobaltate superconductors have not been treated starting from a microscopic SC theory, and no explicit calculations of the doping dependence of the upper critical field have been made so far.

In this case, a challenging issue for theory is to explain the thermodynamic properties of the triangular-lattice cobaltate superconductors. In our recent study \cite{Qin15}, the electromagnetic response in the triangular-lattice cobaltate superconductors is studied based on the kinetic-energy-driven SC mechanism \cite{Feng0306,Feng15}, where we show that the magnetic-field-penetration depth exhibits an exponential temperature dependence due to the absence of the d-wave gap nodes at the charge-carrier Fermi surface. Moreover, in analogy to the dome-shaped doping dependence of $T_{c}$, the superfluid density increases with increasing doping in the lower doped regime, and reaches a maximum around the critical doping, then decreases in the higher doped regime. In this paper, we start from the theoretical framework of the kinetic-energy-driven superconductivity, and then provide a natural explanation to the thermodynamic properties in the triangular-lattice cobaltate superconductors. We evaluate explicitly the internal energy, and then qualitatively reproduced some main features of the heat-capacity and magnetization measurements on the triangular-lattice cobaltate superconductors  \cite{Jin03,Ueland04,Lorenz04,Yang05,Jin05,Sasaki04,Maska04,Chou04,Oeschler08,Oeschler08b}. In particular, we show that a sharp peak in the specific-heat of the triangular-lattice cobaltate superconductors appears at $T_{\rm c}$, and then the specific-heat varies exponentially as a function of temperature for the temperatures $T<T_{\rm c}$ due to the absence of the d-wave gap nodes at the charge-carrier Fermi surface, which is much different from that in the square-lattice cuprate superconductors \cite{Zhao12}, where the characteristic feature is the existence of the gap nodes on the charge-carrier Fermi surface, and then the specific-heat in the square-lattice cuprate superconductors decreases with decreasing temperatures as some power of the temperature in the temperature range $T<T_{\rm c}$. Moreover, the upper critical field follows qualitatively the BCS type temperature dependence, and has the same dome-shaped doping dependence as $T_{c}$.

The rest of this paper is organized as follows. We present the basic formalism in section \ref{framework}, and then the quantitative characteristics of the thermodynamic properties in the triangular-lattice cobaltate superconductors are discussed in section \ref{thermodynamic}, where we show that although the pairing mechanism is driven by the kinetic energy by the exchange of spin excitations \cite{Qin15,Feng0306,Feng15}, the sharp peak of the specific-heat in the triangular-lattice cobaltate superconductors at $T_{\rm c}$ can be described qualitatively by the kinetic-energy-driven d-wave BCS-like formalism. Finally, we give a summary in section \ref{conclusions}.

\section{Formalism}\label{framework}

In the triangular-lattice cobaltate superconductors, the characteristic feature is the presence of the two-dimensional CoO$_{2}$ plane \cite{Takada03,Schaak03,Milne04,Sakurai06,Michioka06}. In this case, a useful microscopic model that has been widely used to describe the low-energy physics of the doped  CoO$_{2}$ plane is the $t$-$J$ model on a triangular lattice \cite{Baskaran03}. This $t$-$J$ model is defined through only two competing parameters: the nearest-neighbor (NN) hopping integral $t$ in the kinetic-energy term, which measures the electron delocalization through the lattice, and the NN spin-spin antiferromagnetic (AF) exchange coupling $J$ in the magnetic-energy part, which describes AF coupling between localized spins. In particular, the NN hopping integral $t$ is much larger than the AF exchange coupling constant $J$ in the Heisenberg term, and therefore the spin configuration is strongly rearranged due to the effect of the charge-carrier hopping $t$ on the spins, which leads to a strong coupling between the charge and spin degrees of freedom of the electron. Since the triangular-lattice cobaltate superconductors are viewed as an electron-doped Mott insulators \cite{Takada03,Schaak03,Milne04,Sakurai06,Michioka06}, this $t$-$J$ model is subject to an important local constraint $\sum_{\sigma}C^{\dagger}_{l\sigma}C_{l\sigma}\geq 1$ to avoid zero occupancy, where $C^{\dagger}_{l\sigma}$ ($C_{l\sigma}$) is the electron creation (annihilation) operator. In the hole-doped side, the local constraint of no double electron occupancy has been treated properly within the fermion-spin approach \cite{Feng15,Feng0494}. However, for an application of the fermion-spin theory to the electron-doped case, we \cite{Qin15,Liu05} should make a particle-hole transformation $C_{l\sigma}\rightarrow f^{\dagger}_{l-\sigma}$, where $f^{\dagger}_{l\sigma}$ ($f_{l\sigma}$) is the hole creation (annihilation) operator, and then the local constraint $\sum_{\sigma}C^{\dagger}_{l\sigma}C_{l\sigma}\geq 1$ without zero occupancy in the electron-doped case is replaced by the local constraint of no double occupancy $\sum_{\sigma} f^{\dagger}_{l\sigma}f_{l\sigma}\leq 1$ in the hole representation. This local constraint of no double occupancy now can be dealt by the fermion-spin theory \cite{Feng15,Feng0494}, where the hole operators $f_{l\uparrow}$ and $f_{l\downarrow}$ are decoupled as $f_{l\uparrow}=a^{\dagger}_{l\uparrow} S^{-}_{l}$ and $f_{l\downarrow}=a^{\dagger}_{l\downarrow}S^{+}_{l}$, respectively, with the charge degree of freedom of the hole together with some effects of spin configuration rearrangements due to the presence of the doped charge carrier itself that are represented by the spinful fermion operator $a_{l\sigma}=e^{-i\Phi_{l\sigma}}a_{l}$, while the spin degree of freedom of the hole is represented by the spin operator $S_{l}$. The advantage of this fermion-spin approach is that the local constraint of no double occupancy is always satisfied in actual calculations.

Based on the $t$-$J$ model in the fermion-spin representation, the kinetic-energy-driven SC mechanism has been developed for the square-lattice cuprate superconductors in the doped regime without an AF long-range order (AFLRO) \cite{Feng0306,Feng15}, where the attractive interaction between charge carriers originates directly from the interaction between charge carriers and spins in the kinetic energy of the $t$-$J$ model by the exchange of spin excitations in the higher powers of the doping concentration. This attractive interaction leads to the formation of the charge-carrier pairs with the d-wave symmetry, while the electron Cooper pairs originated from the charge-carrier d-wave pairing state are due to the charge-spin recombination \cite{Feng15a}, and they condense into the d-wave SC-state. Furthermore, within the framework of the kinetic-energy-driven superconductivity, the doping dependence of the thermodynamic properties in the square-lattice cuprate superconductors has been studied \cite{Zhao12}, and then the striking behavior of the specific-heat in the square-lattice cuprate superconductors are well reproduced. The triangular-lattice cobaltate superconductors on the other hand are the second known example of superconductivity arising from doping a Mott insulator after the square-lattice cuprate superconductors. Although $T_{\rm c}$ in the triangular-lattice cobaltate superconductors is much less than that in the square-lattice cuprate superconductors, the strong electron correlation is common for both these materials, which suggest that these two oxide systems may have the same underlying SC mechanism. In this case, the kinetic-energy-driven superconductivity developed for the square-lattice cuprate superconductors has been generalized to the case for the triangular-lattice cobaltate superconductors \cite{Qin15,Liu05}. The present work of the discussions of the thermodynamic properties in the triangular-lattice cobaltate superconductors builds on the kinetic-energy-driven SC mechanism developed in Refs. \cite{Qin15} and \cite{Liu05}, and only a short summary of the formalism is therefore given in the following discussions. In our previous discussions in the doped regime without AFLRO, the full charge-carrier diagonal and off-diagonal Green's functions of the $t$-$J$ model on a triangular lattice in the charge-carrier pairing state have been obtained explicitly as \cite{Qin15,Liu05},
\begin{subequations}\label{BCSGF}
\begin{eqnarray}
g({\bf k},\omega)&=&Z_{\rm aF}\left ({U^{2}_{{\rm a}{\bf k}}\over\omega-E_{{\rm a}{\bf k}}}+{V^{2}_{{\rm a}{\bf k}}\over\omega+E_{{\rm a}{\bf k}}}\right ), \label{BCSDGF}\\
\Gamma^{\dagger}({\bf k},\omega)&=&-Z_{\rm aF}{\bar{\Delta}^{({\rm a})}_{{\rm Z}{\bf k}}\over 2E_{{\rm a}{\bf k}}}\left ({1\over \omega-E_{{\rm a}{\bf k}}}-{1\over\omega
+E_{{\rm a}{\bf k}}}\right ),~~~~~\label{BCSODGF}
\end{eqnarray}
\end{subequations}
where the charge-carrier quasiparticle coherent weight $Z_{\rm aF}$, the charge-carrier quasiparticle coherence factors $2U^{2}_{{\rm a}{\bf k}}=1+\bar{\xi}_{{\bf k}}/E_{{\rm a}{\bf k}}$ and $2V^{2}_{{\rm a}{\bf k}}=1-\bar{\xi}_{{\bf k}}/E_{{\rm a}{\bf k}}$, the charge-carrier quasiparticle energy spectrum $E_{{\rm a}{\bf k}}=\sqrt{\bar{\xi}^{2}_{\bf k}+ \mid \bar{\Delta}^{({\rm a})}_{{\rm Z}{\bf k}}\mid^{2}}$, and the charge-carrier excitation spectrum $\bar{\xi}_{{\bf k}}$. In the early days of superconductivity in the triangular-lattice cobaltate superconductors, some NMR and NQR data are consistent with the case of the existence of a pair gap over the Fermi surface \cite{Kobayashi03,Michioka06}, while other experimental NMR and NQR results suggest the existence of the gap nodes \cite{Fujimoto04,Zheng06}. In particular, it has been argued that only involving the pairings of charge carriers located at the next NN sites can give rise to the nodal points of the complex gap appearing inside the Brillouin zone \cite{Valkov15,Zhou08}. Moreover, the nodal points of the complex gap has been obtained theoretically by considering the interaction between the Hubbard fermions \cite{Valkov15}. However, although the recent experimental results \cite{Oeschler08} obtained from the specific-heat measurements do not give unambiguous evidence for either the presence or absence of the nodes in the energy gap, the experimental data of the specific-heat \cite{Oeschler08} are consistent with these fitted results obtained from phenomenological BCS formalism with the d-wave symmetry without gap nodes. Furthermore, some theoretical calculations based on the numerical simulations indicate that the d-wave state without gap nodes is the lowest state around the electron-doped regime where superconductivity appears in triangular-lattice cobaltate superconductors \cite{Honerkamp03,Watanabe04,Weber06}. In particular, the recent theoretical studies based on a large-scale dynamical cluster quantum Monte Carlo simulation \cite{Chen13} and a combined cluster calculation and renormalization group approach \cite{Kiesel13} show that the d-wave state naturally explains some SC-state properties as indicated by experiments. In this case, we only consider the case with the d-wave pairing symmetry as our previous discussions \cite{Qin15}, and then the d-wave charge-carrier pair gap $\bar{\Delta}^{({\rm a})}_{{\rm Z}{\bf k}}$ in Eq. (\ref{BCSGF}) has been given in Ref. \cite{Qin15}.

Since the spin part in the $t$-$J$ model in the fermion-spin representation is anisotropic away from half-filling \cite{Feng15}, two spin Green's functions $D(l-l',t-t')=\langle\langle S^{+}_{l}(t);S^{-}_{l'}(t')\rangle\rangle$ and $D_{\rm z}(l-l',t-t')=\langle\langle S^{\rm z}_{l}(t);S^{\rm z}_{l'}(t')\rangle\rangle$ have been defined to describe properly the spin part, and can be obtained explicitly as,
\begin{subequations}\label{SGF}
\begin{eqnarray}
D({\bf k},\omega)&=&{B_{\bf k}\over 2\omega_{\bf k}}\left ({1\over \omega-\omega_{\bf k}}-{1\over\omega+\omega_{\bf k}}\right ), \label{MFSGF}\\
D_{\rm z}({\bf k},\omega)&=& {B_{{\rm z}{\bf k}}\over 2\omega_{{\rm z}{\bf k}}}\left ({1\over\omega-\omega_{{\rm z}{\bf k}}}-{1\over \omega+\omega_{{\rm z}{\bf k}}}\right ), \label{MFSGFZ}
\end{eqnarray}
\end{subequations}
where the function $B_{\bf k}$ and spin excitation spectrum $\omega_{\bf k}$ in the spin Green's function (\ref{MFSGF}) have been given in Ref. \cite{Qin15}, while the function $B_{{\rm z}{\bf k}}=\epsilon\lambda\chi(\gamma_{\bf k}-1)$, and the spin excitation spectrum $\omega_{{\rm z}{\bf k}}$ in the spin Green's function (\ref{MFSGFZ}) is obtained as,
\begin{eqnarray}
\omega^{2}_{{\rm z}{\bf k}}&=&\epsilon\lambda^{2}\left [\alpha\chi\gamma_{\bf k}-\epsilon A_{1}+{1\over 6}\alpha\chi \right](\gamma_{\bf k} -1),
\end{eqnarray}
where $\gamma_{\bf k}=[\cos k_{x}+2\cos(k_{x}/2)\cos(\sqrt{3}k_{y}/2)]/3$, the parameters $A_{1}$, $\lambda$, $\epsilon$, the decoupling parameter $\alpha$, and the spin correlation function $\chi$ have been also given in Ref. \cite{Qin15}. In particular, the charge-carrier quasiparticle coherent weight $Z_{\rm aF}$, the charge-carrier pair gap parameter $\bar{\Delta}^{({\rm a})}$, all the other order parameters, and the decoupling parameter $\alpha$ have been determined by the self-consistently calculation \cite{Qin15}. In spite of the pairing mechanism driven by the kinetic energy by the exchange of spin excitations, the results in Eq. (\ref{BCSGF}) are the standard BCS expressions for a d-wave charge-carrier pair state.

Now we turn to evaluate the the internal energy of the triangular-lattice cobaltate superconductors. The internal energy in the charge-spin separation fermion-spin representation can be expressed as \cite{Zhao12} $U_{\rm total}(T)=U_{\rm charge}(T)+U_{\rm spin}(T)$, where $U_{\rm charge}(T)$ and $U_{\rm spin}(T)$ are the corresponding contributions from charge carriers and spins, respectively, and can be obtained in terms of the charge-carrier spectral function $A_{\rm charge}({\bf k},\omega,T)=-2{\rm Im}g({\bf k},\omega)$, and the spin spectral functions $A_{\rm spin}({\bf k},\omega, T)=-2{\rm Im}D({\bf k},\omega)$ and $A^{(\rm z)}_{\rm spin}({\bf k},\omega,T)=-2{\rm Im}D_{\rm z}({\bf k},\omega)$. Following the previous work for the case in the square-lattice cuprate superconductors \cite{Zhao12}, it is straightforward to find the internal energy of the triangular-lattice cobaltate superconductors in the SC-state as,
\begin{eqnarray}\label{ES}
U^{(\rm s)}_{\rm total}(T)&=&-{Z_{\rm aF}\over N}\sum_{\bf k}[E_{{\rm a}{\bf k}}{\rm tanh}({1\over 2}\beta E_{{\rm a}{\bf k}})]\nonumber\\
&+&{Z_{\rm aF}\over N}\sum_{\bf k}\bar{\xi}_{\bf k} +6J_{\rm eff}(\chi+\chi^{\rm z}),
\end{eqnarray}
where $J_{\rm eff}=(1-\delta)^{2}J$ with the doping concentration $\delta$, while the spin correlation function $\chi^{\rm z}$ has been given in Ref. \cite{Qin15}. In the normal-state, the charge carrier pair gap $\bar{\Delta}^{(\rm a)}=0$, and in this case, the SC-state internal energy (\ref{ES}) can be reduced to the normal-state case as,
\begin{eqnarray}\label{EN}
U^{(\rm n)}_{\rm total}(T)&=&-{Z_{\rm aF}\over N}\sum_{\bf k}[\bar{\xi}_{\bf k}{\rm tanh}({1\over 2}\beta\bar{{\xi}_{\bf k}})]\nonumber\\
&+&{Z_{\rm aF}\over N}\sum_{\bf k}\bar{\xi}_{\bf k} +6J_{\rm eff}(\chi+\chi^{\rm z}).
\end{eqnarray}

\section{Thermodynamic properties}\label{thermodynamic}

\begin{figure}[h!]
\centering
\includegraphics[scale=0.28]{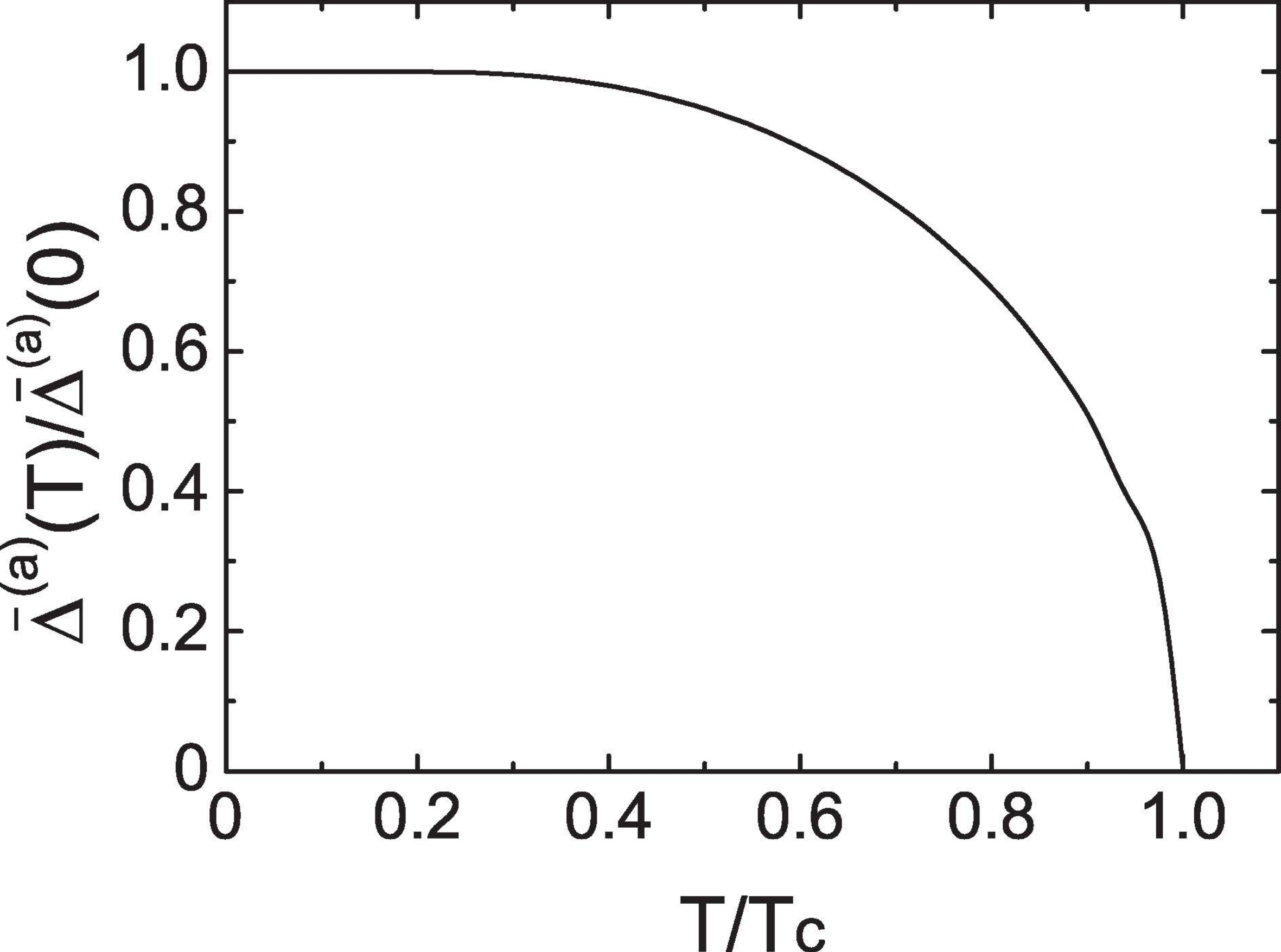}
\caption{The charge-carrier pair gap parameter as a function of temperature at $\delta=0.15$ for $t/J=-2.5$. \label{pair-gap-parameter-temp}}
\end{figure}

We are now ready to discuss the thermodynamic properties in the triangular-lattice cobaltate superconductors. The charge-carrier pair gap parameter $\bar{\Delta}^{({\rm a})}$ is one of the characteristic parameters in the triangular-lattice cobaltate superconductors, which incorporates both the pairing force and charge-carrier pair order parameter, and therefore measures the strength of the binding of two charge carriers into a charge-carrier pair. In particular, the charge-carrier pair order parameter and the charge-carrier pair macroscopic wave functions in the triangular-lattice cobaltate superconductors are the same within the framework of the kinetic-energy-driven SC mechanism \cite{Feng15}, i.e., the charge-carrier pair order parameter is a {\it magnified} version of the charge-carrier pair macroscopic wave functions. For the convenience in the following discussions, we plot the charge-carrier pair gap parameter $\bar{\Delta}^{({\rm a})}$ as a function of temperature at the doping concentration $\delta=0.15$ for parameter $t/J=-2.5$ in Fig. \ref{pair-gap-parameter-temp}. It is shown clearly that the charge-carrier pair gap parameter follows qualitatively a BCS-type temperature dependence, i.e., it decreases with increasing temperatures, and eventually vanishes at $T_{\rm c}$.

\subsection{Specific-heat}

\begin{figure}[h!]
\centering
\includegraphics[scale=0.28]{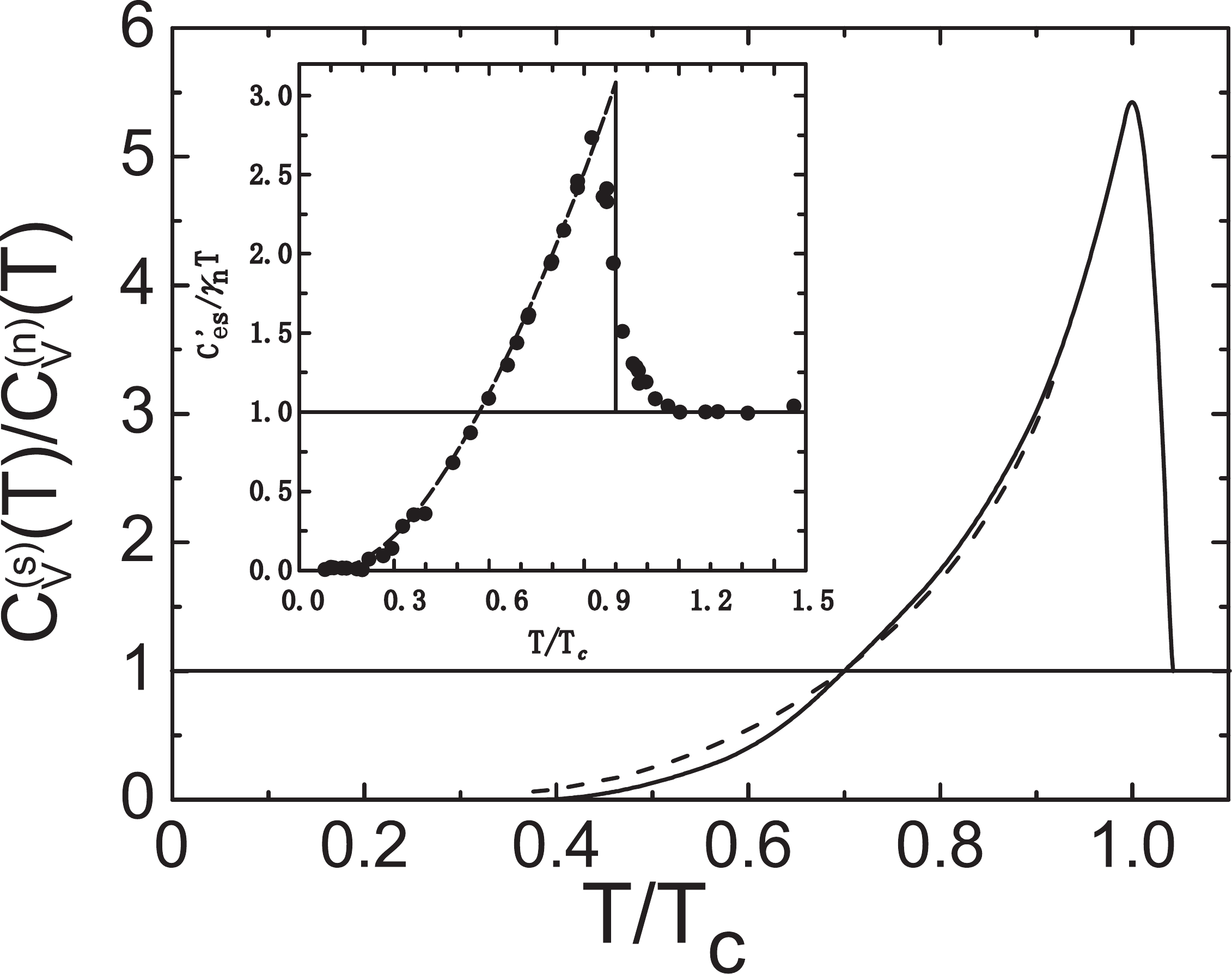}
\caption{The specific-heat as a function of temperature at $\delta=0.20$ for $t/J=-2.5$ and $J=50$meV. The dashed line is obtained from a numerical fit $C^{(\rm s)}_{\rm v}/C^{(\rm n)}_{\rm v}= A\exp[-B\bar{\Delta}^{({\rm a})}(T)/T]$, with $A\sim 10.69$ and $B\sim 270.65$. Inset: the corresponding experimental data of Na$_{x}$CoO$_{2}$$\cdot y$H$_{2}$O taken from Ref. \cite{Oeschler08}. \label{specific-heat}}
\end{figure}

One of the characteristics quantites in the thermodynamic properties is the specific-heat, which can be obtained by evaluating the temperature-derivative of the internal energy as,
\begin{subequations}\label{heat}
\begin{eqnarray}
C^{(\rm s)}_{\rm v}(T)={{\rm d}U^{(\rm s)}(T)\over {\rm d}T}=\gamma_{\rm s}(T)T, \label{SC-heat}\\
C^{(\rm n)}_{\rm v}(T)={{\rm d}U^{(\rm n)}(T)\over {\rm d}T}=\gamma_{\rm n}(T)T, \label{N-heat}
\end{eqnarray}
\end{subequations}
in the SC-state and normal-state, respectively, where $\gamma_{\rm s}(T)$ and $\gamma_{\rm n}(T)$ are the temperature dependence of the specific-heat coefficients in the SC-state and normal-state, respectively. In Fig. \ref{specific-heat}, we plot the specific-heat $C^{(\rm s)}_{\rm v}/C^{(\rm n)}_{\rm v}$ (solid line) as a function of temperature at $\delta= 0.20$ for $t/J=-2.5$ and $J=50$meV. For comparison, the corresponding experimental result \cite{Oeschler08} of Na$_{x}$CoO$_{2}$$\cdot y$H$_{2}$O is also shown in Fig. \ref{specific-heat} (inset). Apparently, the main feature of the specific-heat observed experimentally on the triangular-lattice cobaltate superconductors \cite{Jin03,Ueland04,Lorenz04,Yang05,Jin05,Sasaki04,Maska04,Chou04,Oeschler08,Oeschler08b} is qualitatively reproduced. As can be seen from Fig. \ref{specific-heat}, the specific-heat anomaly (a jump) at $T_{\rm c}$ appears. The SC transition is reflected by a sharp peak in the specific-heat at $T_{\rm c}$, however, the magnitude of the specific-heat decreases dramatically with decreasing temperatures for the temperatures $T<T_{\rm c}$. Moreover, the calculated result of the specific-heat difference $\Delta C_{\rm v}(T_{\rm c}) /C^{(\rm n)}_{\rm v}(T_{\rm c})=[C^{(\rm s)}_{\rm v}(T_{\rm c})-C^{(\rm n)}_{\rm v}(T_{\rm c})]/C^{(\rm n)}_{\rm v}(T_{\rm c})=4.7$ for the discontinuity in the specific-heat at $T_{\rm c}$, which is roughly consistent with the experimental data \cite{Oeschler08b} $\Delta C_{\rm v}(T_{\rm c})/C^{(\rm n)}_{\rm v}(T_{\rm c})\approx 2.08$ observed on Na$_{x}$CoO$_{2}$$\cdot y$H$_{2}$O. For a better understanding of the physical properties of the specific-heat in the triangular-lattice superconductors, we have fitted our present theoretical result of the specific-heat for the temperatures $T<T_{\rm c}$, and the fitted result is also plotted in Fig. \ref{specific-heat} (dashed line), where we found that $C^{(\rm s)}_{\rm v} /C^{(\rm n)}_{\rm v}$ varies exponentially as a function of temperature ($C^{(\rm s)}_{\rm v}/C^{(\rm n)}_{\rm v}= A\exp[-B\bar{\Delta}^{({\rm a})}(T)/T]$ with $A\sim 10.69$ and $B\sim 270.65$), which is an expected result in the case without the d-wave gap nodes at the charge-carrier Fermi surface, and is in qualitative agreement with experimental data \cite{Jin03,Ueland04,Lorenz04,Yang05,Jin05,Sasaki04,Maska04,Chou04,Oeschler08,Oeschler08b}. However, this result in the triangular-lattice superconductors is much different from that in the square-lattice cuprate superconductors, where the characteristic feature is the existence of the gap nodes on the charge-carrier Fermi surface, and then the specific-heat of the square-lattice cuprate superconductors decreases with decreasing temperatures as some power of the temperature for the temperatures $T<T_{\rm c}$.

\subsection{Condensation energy}

\begin{figure}[h!]
\centering
\includegraphics[scale=0.28]{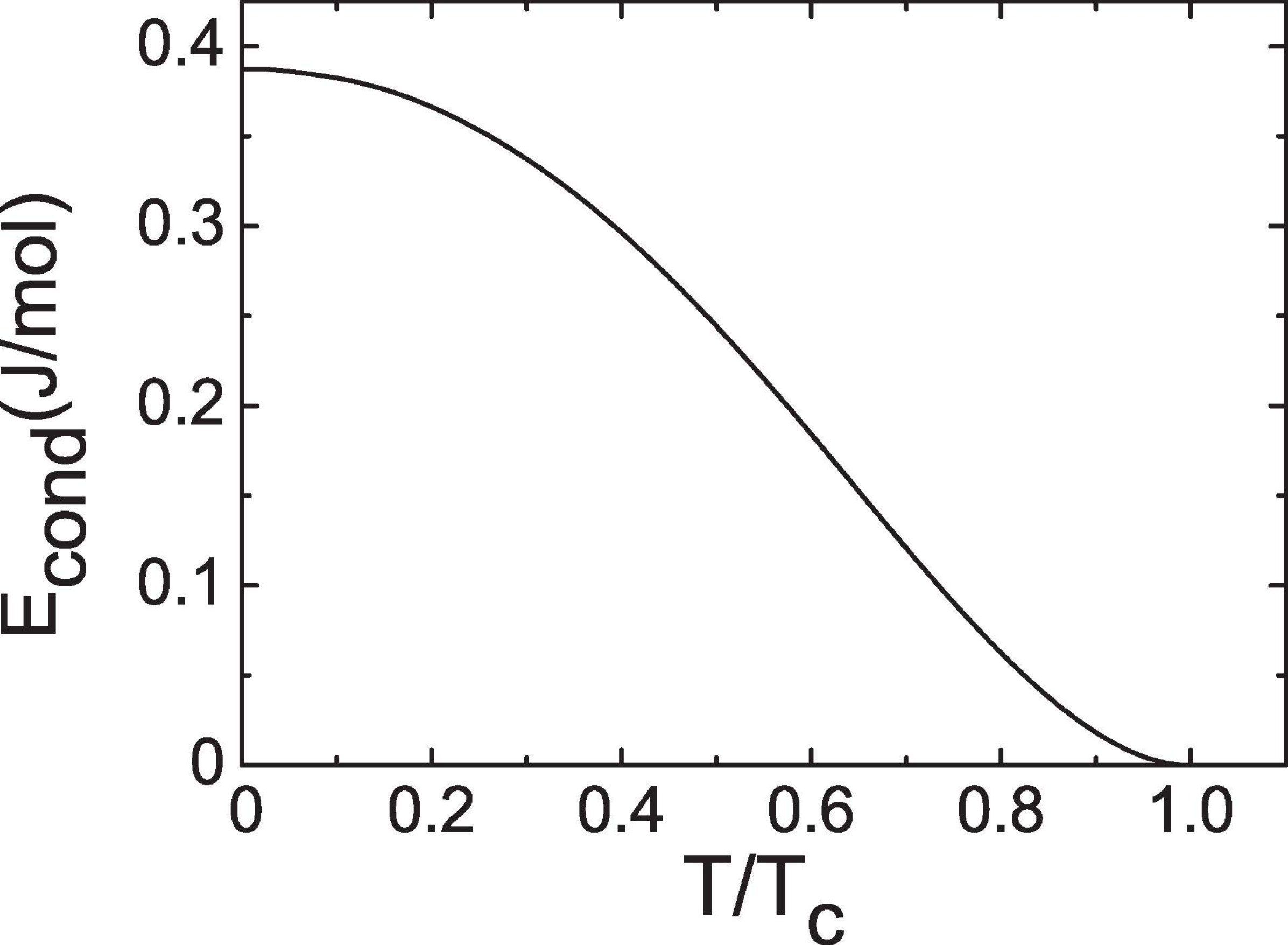}
\caption{The condensation energy as a function of temperature at $\delta=0.20$ for $t/J=-2.5$ and $J=50$meV. \label{Econd}}
\end{figure}

In the framework of the kinetic-energy-driven SC mechanism \cite{Qin15,Liu05}, the exchanged bosons are spin excitations that act like a bosonic glue to hold the charge-carrier pairs together, and then these charge-carrier pairs (then electron pairs) condense into the SC-state. As a consequence, the charge-carrier pairs in the triangular-lattice cobaltate superconductors are always related to lower the total free energy. The condensation energy $E_{\rm cond}(T)$ on the other hand is defined as the energy difference between the normal-state free energy, extrapolated to zero temperature, and the SC-state free energy,
\begin{eqnarray}\label{condensation-energy}
E_{\rm cond}(T)&=&[U^{(\rm n)}(T)-TS^{(\rm n)}(T)]\nonumber\\
&-&[U^{(\rm s)}(T)-TS^{(\rm s)}(T)],
\end{eqnarray}
where the related entropy of the system is evaluated from the specific-heat coefficient in Eq. (\ref{heat}) as,
\begin{eqnarray}\label{entropy}
S^{(\rm a)}(T)=\int\limits_{0}^{T}\gamma_{\rm a}(T'){\rm d}T',
\end{eqnarray}
where ${\rm a}={\rm s}$, ${\rm n}$ referring to the SC-state and normal-state, respectively. We have made a calculation for the condensation energy (\ref{condensation-energy}), and the result of $E_{\rm cond}(T)$ as a function of temperature at $\delta=0.20$ for $t/J=-2.5$ and $J=50$meV is plotted in Fig. \ref{Econd}. In comparison with the result of the temperature dependence of the charge-carrier pair gap parameter shown in Fig. \ref{pair-gap-parameter-temp}, we therefore find that in spite of the pairing mechanism driven by the kinetic energy by the exchange of spin excitations, the condensation energy of the triangular-lattice cobaltate superconductors follows qualitatively a BCS type temperature dependence.

\subsection{Upper critical field}

\begin{figure}[h!]
\centering
\includegraphics[scale=0.28]{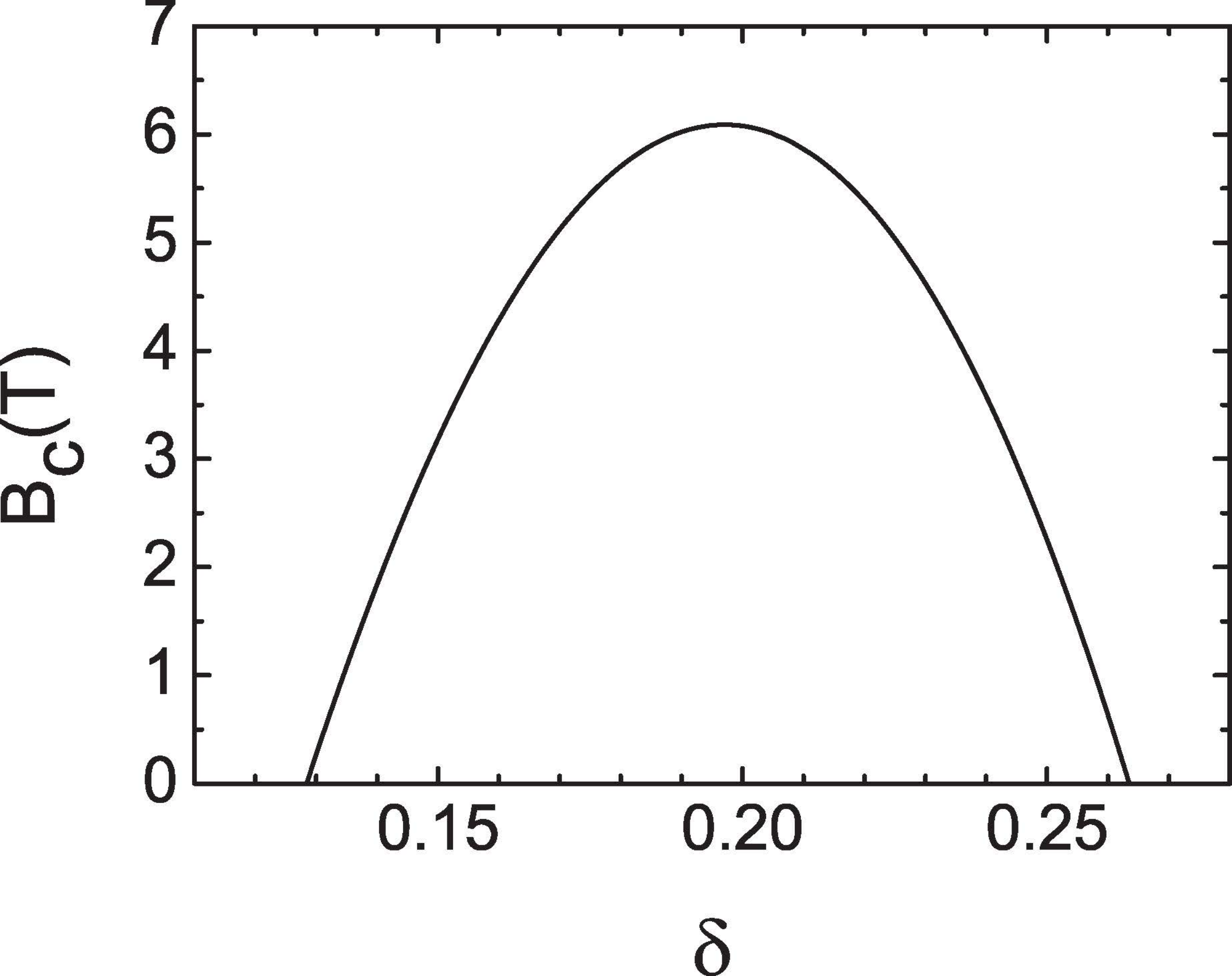}
\caption{The upper critical field as a function of doping with $T=0.0002J$ for $t/J=-2.5$ and $J=50$meV. \label{Bc-doping}}
\end{figure}

\begin{figure}[h!]
\centering
\includegraphics[scale=0.35]{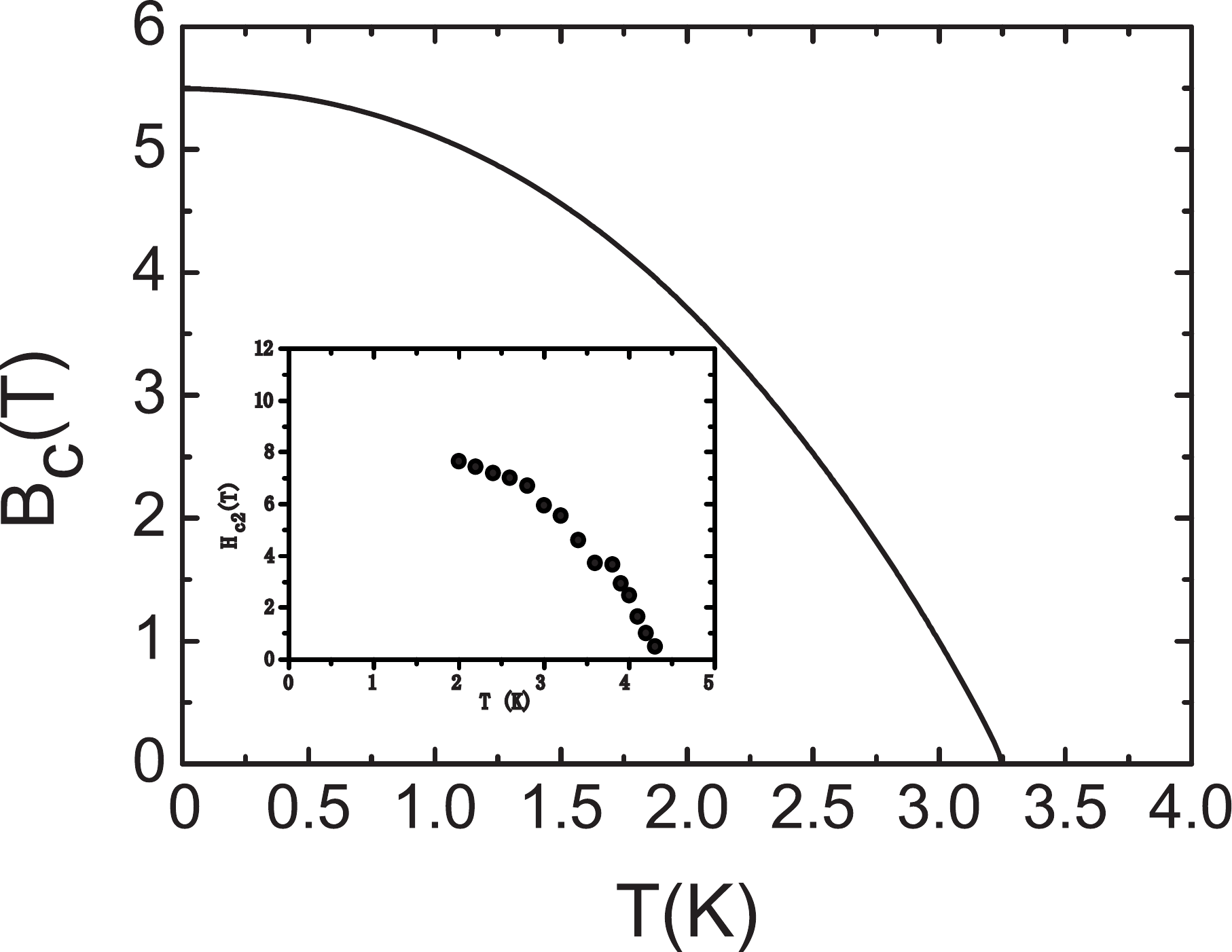}
\caption{The upper critical field as a function of temperature at $\delta=0.20$ for $t/J=-2.5$ and $J=50$meV. Insets: the corresponding experimental data of Na$_{x}$CoO$_{2}$$\cdot y$H$_{2}$O taken from Ref. \cite{Chou04}. \label{Bc-temp}}
\end{figure}

A quantity which is directly related to the condensation energy $E_{\rm cond}(T)$ in Eq. (\ref{condensation-energy}) is the upper critical field $B_{\rm c}(T)$,
\begin{eqnarray}\label{field}
{1\over 2\mu_{0}}B^{2}_{\rm c}(T)=E_{\rm cond}(T).
\end{eqnarray}
This upper critical field $B_{\rm c}(T)$ is a fundamental parameter whose variation as a function of doping and temperature provides important information crucial to understanding the details of the SC-state. In Fig. \ref{Bc-doping}, we plot the upper critical field $B_{\rm c}(T)$ as a function of doping with $T=0.0002J$ for $t/J=-2.5$ and $J=50$meV. It is shown clearly that the upper critical field takes a dome-shaped doping dependence with the underdoped and overdoped regimes on each side of the optimal doping, where  $B_{\rm c}(T)$ reaches its maximum. Moreover, the calculated upper critical field at the optimal doping is $B_{\rm c}\approx 6.1T$, which is not too far from the range $B_{\rm c}\approx 1.7T\sim 9T$ estimated experimentally for different samples of Na$_{x}$CoO$_{2}\cdot y$H$_{2}$O \cite{Jin05,Sasaki04,Maska04,Chou04}. For a superconductor, the upper critical field is defined as the critical magnetic field that destroys the SC-state at zero temperature, which therefore means that the upper critical field also measures the strength of the binding of charge carriers into the charge-carrier pairs. In this case, the domelike shape of the doping dependence of $B_{\rm c}(T)$ is a natural consequence of the domelike shape of the doping dependence of $\bar{\Delta}^{({\rm a})}$ and $T_{\rm c}$ as shown in Ref. \cite{Qin15}. To further understand the intrinsic property of the upper critical field $B_{\rm c}(T)$ in the triangular-lattice cobaltate superconductors, we have also performed a calculation for $B_{\rm c}(T)$ at different temperatures, and the result of $B_{\rm c}(T)$ as a function of temperature at $\delta=0.20$ for $t/J=-2.5$ and $J=50$meV is plotted in Fig. \ref{Bc-temp} in comparison with the corresponding experimental result \cite{Chou04} of Na$_{x}$CoO$_{2}\cdot y$H$_{2}$O (inset). It is thus shown that $B_{\rm c}(T)$ varies moderately with initial slope. In particular, as in the case of the temperature dependence of the condensation energy shown in Fig. \ref{Econd}, the upper critical field $B_{\rm c}(T)$ also follows qualitatively the BCS type temperature dependence, i.e., it decreases with increasing temperature, and vanishes at $T_{\rm c}$, which is also qualitatively consistent with the experimental results \cite{Jin05,Sasaki04,Maska04,Chou04}.

\begin{figure}[h!]
\centering
\includegraphics[scale=0.28]{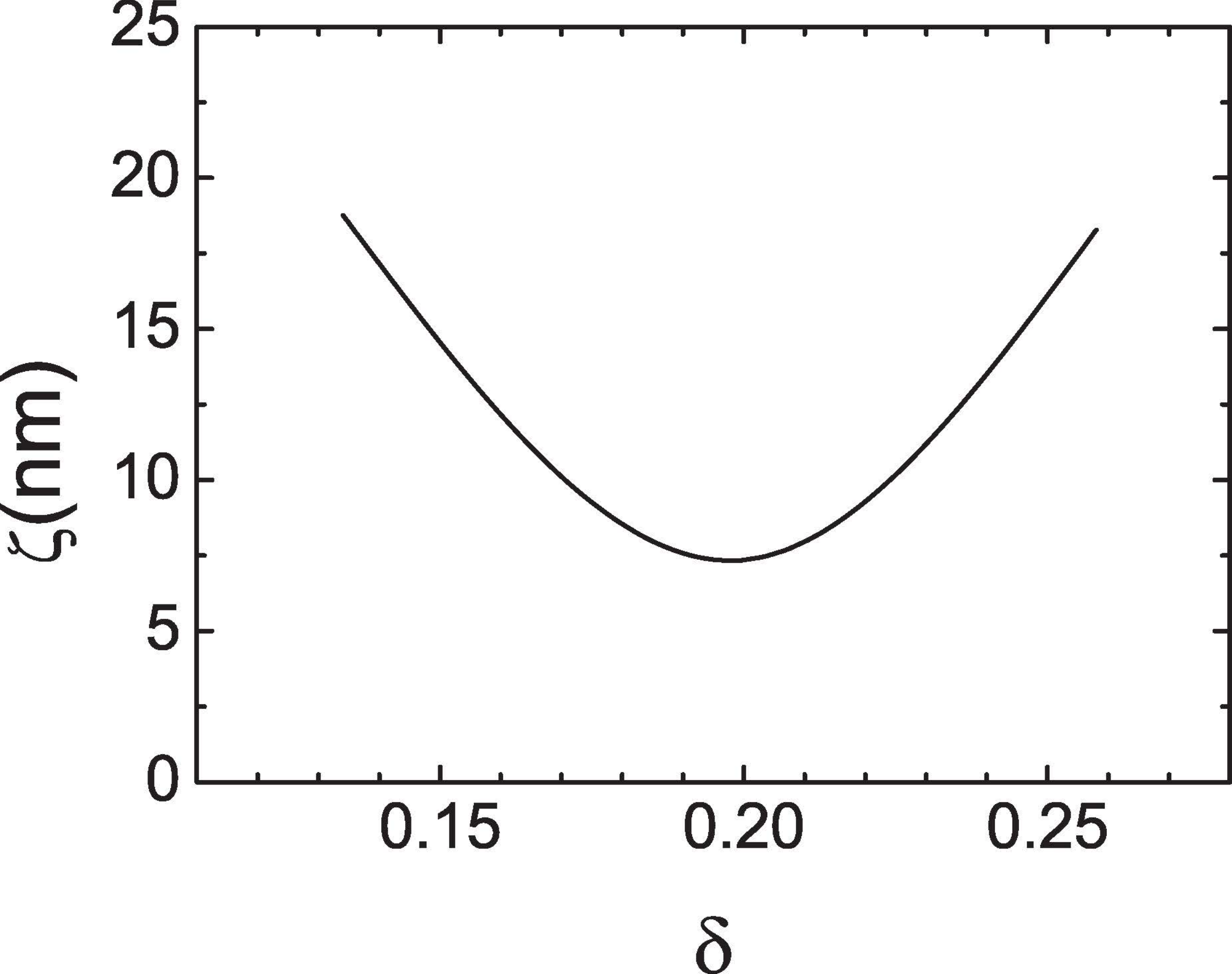}
\caption{The coherence length as a function of doping at $T=0.0002J$ for $t/J=-2.5$ and $J=50$meV. \label{coherence}}
\end{figure}

The coherence length $\zeta(T)$ also is one of the basic SC parameters of the triangular-lattice cobaltate superconductors, and is directly associated with the upper critical field as $\zeta^{2}(T)=\Phi_{0}/[2\pi B_{\rm c}(T)]$, where $\Phi_{0}=hc/(2e)$ is the magnetic flux quantum. In Fig. \ref{coherence}, we plot the coherence length $\zeta(T)$ as a function of doping with $T=0.0002J$ for $t/J=-2.5$ and $J=50$meV. Since the coherence length $\zeta(T)$ is inversely proportional to the upper critical field $B_{\rm c}(T)$, the coherence length $\zeta(T)$ in the triangular-lattice cobaltate superconductors reaches a minimum around the optimal doping, then grows in both the underdoped and overdoped regimes. In particular, at the optimal doping, the anticipated coherence length $\zeta_{\rm opt}\approx 7.1$nm approximately matches the coherence length $\zeta_{\rm opt}\approx 4.4$nm observed in the optimally doped Na$_{x}$CoO$_{2}$$\cdot y$H$_{2}$O \cite{Jin05}. This coherence length $\zeta_{\rm opt}\approx 7.1$nm at the optimal doping estimated from the upper critical field using the Ginzburg-Landau expression also is qualitatively consistent with that obtained based on the microscopic calculation $\zeta_{\rm opt}=\hbar v_{\rm F}/(\pi\Delta^{({\rm a})})\approx 7.94$nm, where $v_{\rm F} =\hbar^{-1} \partial\xi_{\bf k}/\partial {\bf k}|_{k_{\bf F}}$ is the charge carrier velocity at the Fermi surface. This relatively short coherence length is surprising for a superconductor with such a low $T_{\rm c}$, but is consistent with the narrow bandwidth in the triangular-lattice superconductors \cite{Chou04}, since the charge-carrier quasiparticle spectrum $E_{{\rm a}{\bf k}}=\sqrt{\bar{\xi}^{2}_{\bf k} +\mid\bar{\Delta}_{\rm aZ} ({\bf k})\mid^{2}}$ in the full charge-carrier diagonal Green's function (\ref{BCSDGF}) and off-diagonal Green's function (\ref{BCSODGF}) has a narrow bandwidth $W_{\rm a}\sim 2J$.

\section{Conclusions}\label{conclusions}

Within the framework of the kinetic-energy-driven SC mechanism, we have discussed the doping dependence of the thermodynamic properties in the triangular-lattice cobaltate superconductors. We show that the specific-heat anomaly (a jump) appears at $T_{\rm c}$, and then the specific-heat varies exponentially as a function of temperature for the temperatures $T<T_{\rm c}$ due to the absence of the d-wave gap nodes at the charge-carrier Fermi surface, which is much different from that in the square-lattice cuprate superconductors \cite{Zhao12}, where the characteristic feature is the existence of the gap nodes on the charge-carrier Fermi surface, and then the specific-heat of the square-lattice cuprate superconductors decreases with decreasing temperatures as some power of the temperature in the temperature range $T<T_{\rm c}$. On the other hand, both the condensation energy and the upper critical field in the triangular-lattice cobaltate superconductors follow qualitatively the BCS type temperature dependence. In particular, in analogy to the dome-shaped doping dependence of $T_{\rm c}$, the maximal upper critical field occurs around the optimal doping, and then decreases in both underdoped and overdoped regimes. Incorporating the present result \cite{Zhao12} with that obtained in the square-lattice cuprate superconductors, it is thus shown that the dome-shaped doping dependence of the upper critical field is a universal feature in a doped Mott insulator, and it does not depend on the details of the geometrical spin frustration. Since the knowledge of the thermodynamic properties in the triangular-lattice cobaltate superconductors is of considerable importance as a test for theories of superconductivity, the qualitative agreement between the present theoretical results and experimental data also provides an important confirmation of the nature of the SC phase of the triangular-lattice cobaltate superconductors as a conventional BCS-like with the d-wave symmetry, although the pair mechanism is driven by the kinetic energy by the exchange of spin excitations.


\begin{thebibliography}{00}

\bibitem{Bednorz86} J. G. Bednorz and K. A. M\"uller, Z. Phys. B {\bf 64}, 189 (1986).

\bibitem{Kastner98} See, e.g., the review, M. A. Kastner, R. J. Birgeneau, G. Shirane, and Y. Endoh, Rev. Mod. Phys. {\bf 70}, 897 (1998).

\bibitem{Tallon95} J. L. Tallon, C. Bernhard, H. Shaked, R. L. Hitterman, and J. D. Jorgensen, Phys. Rev. B {\bf 51}, 12911 (1995).

\bibitem{Takada03} Kazunori Takada, Hiroya Sakurai, Eiji Takayama-Muromachi, Fujio Izumi, Ruben A. Dilanian, and Takayoshi Sasaki, Nature {\bf 422}, 53 (2003).

\bibitem{Schaak03} R. E. Schaak, T. Klimczuk, M. L. Foo, and R. J. Cava, Nature {\bf 424}, 527 (2003).

\bibitem{Milne04} C. J. Milne, D. N. Argyriou, A. Chemseddine, N. Aliouane, J. Veira, S. Landsgesell, and D. Alber, Phys. Rev. Lett. {\bf 93}, 247007 (2004).

\bibitem{Sakurai06} Hiroya Sakurai, Naohito Tsujii, Osamu Suzuki, Hideaki Kitazawa, Giyuu Kido, Kazunori Takada, Takayoshi Sasaki, and Eiji Takayama-Muromachi, Phys. Rev. B {\bf 74}, 092502 (2006).

\bibitem{Michioka06} C. Michioka, H. Ohta, Y. Itoh, and K. Yoshimura, J. Phys. Soc. Jpn. {\bf 75}, 063701 (2006).

\bibitem{Schrieffer64} J. R. Schrieffer, {\it Theory of Superconductivity}, Benjamin, New York, 1964.

\bibitem{Jin03} R. Jin, B. C. Sales, P. Khalifah, and D. Mandrus, Phys. Rev. Lett. {\bf 91}, 217001 (2003).

\bibitem{Ueland04} B.G. Ueland, P. Schiffer, R.E. Schaak, M.L. Foo, V.L. Miller, and R.J. Cava, Physica C {\bf 402}, 27 (2004).

\bibitem{Lorenz04} B. Lorenz, J. Cmaidalka, R.L. Meng, and C.W. Chu, Physica C {\bf 402}, 106 (2004).

\bibitem{Yang05} H. D. Yang, J.-Y. Lin, C. P. Sun, Y. C. Kang, C. L. Huang, K. Takada, T. Sasaki, H. Sakurai, and E. Takayama-Muromachi, Phys. Rev. B {\bf 71}, 020504(R) (2005).

\bibitem{Jin05} R. Jin, B. C. Sales, S. Li, and D. Mandrus, Phys. Rev. B {\bf 72}, 060512(R) (2005).

\bibitem{Sasaki04} Takahiko Sasaki, Petre Badica, Naoki Yoneyama, Kazuyoshi Yamada, Kazumasa Togano, and Norio Kobayashi, J. Phys. Soc. Jpn. {\bf 73}, 1131 (2004).

\bibitem{Maska04} M. M. Ma\'ska, M. Mierzejewski, B. Andrzejewski, M. L. Foo, R. J. Cava, and T. Klimczuk, Phys. Rev. B {\bf 70}, 144516 (2004).

\bibitem{Chou04} F. C. Chou, J. H. Cho, P. A. Lee, E. T. Abel, K. Matan, and Y. S. Lee, Phys. Rev. Lett. {\bf 92}, 157004 (2004).

\bibitem{Oeschler08} N. Oeschler, R. A. Fisher, N. E. Phillips, J. E. Gordon, M.-L. Foo and R. J. Cava, EPL {\bf 82}, 47011 (2008).

\bibitem{Oeschler08b} N. Oeschler, R. A. Fisher, N. E. Phillips, J. E. Gordon, M.-L. Foo, and R. J. Cava, Phys. Rev. B {\bf 78}, 054528 (2008).

\bibitem{Kobayashi03} Yoshiaki Kobayashi, Mai Yokoi, Masatoshi Sato, J. Phys. Soc Jpn. {\bf 72}, 2453 (2003).

\bibitem{Fujimoto04} Tatsuya Fujimoto, Guo-qing Zheng, Y. Kitaoka, R. L. Meng, J. Cmaidalka, and C. W. Chu, Phys. Rev. Lett. {\bf 92}, 047004 (2004).

\bibitem{Ihara06} Yoshihiko Ihara, Kenji Ishida, Hideo Takeya, Chishiro Michioka, Masaki Kato, Yutaka Itoh, Kazuyoshi Yoshimura, Kazunori Takada, Takayoshi Sasaki, Hiroya Sakurai, and Eiji Takayama-Muromachi, J. Phys. Soc. Jpn. {\bf 75}, 013708 (2006).

\bibitem{Kato06} M. Kato, C. Michioka, T. Waki, Y. Itoh, K. Yoshimura, K. Ishida, H. Sakurai, E. Takayama-Muromachi, K. Takada, and T. Sasaki, J. Phys.: Condens. Matter {\bf 18}, 669 (2006).

\bibitem{Zheng06} Guo-qing Zheng, Kazuaki Matano, R. L. Meng, J. Cmaidalka, and C. W. Chu, J. Phys.: Condens. Matter {\bf 18}, L63 (2006).

\bibitem{Mazin05} I. I. Mazin and M. D. Johannes, Nature Phys. {\bf 1}, 91 (2005).

\bibitem{Lee90} T. K. Lee and Shiping Feng, Phys. Rev. B {\bf 41}, 11110 (1990).

\bibitem{Baskaran03} G. Baskaran, Phys. Rev. Lett. {\bf 91}, 097003 (2003).

\bibitem{Kumar04} Brijesh Kumar and B. Sriram Shastry, Phys. Rev. B {\bf 68}, 104508 (2003).

\bibitem{Wang04} Qiang-Hua Wang, Dung-Hai Lee, and Patrick A. Lee, Phys. Rev. B {\bf 69}, 092504 (2004).

\bibitem{Ogata03} Masao Ogata, J. Phys. Soc. Jpn. {\bf 72}, 1839 (2003).

\bibitem{Liu05} Bin Liu, Ying Liang, Shiping Feng, and Wei-Yeu Chen, Commun. Theor. Phys. {\bf 43}, 1127 (2005).

\bibitem{Qin15} Ling Qin, Xixiao Ma, L\"ulin Kuang, Jihong Qin, and Shiping Feng, J. Low Temp. Phys. {\bf 181}, 112-133 (2015).

\bibitem{Chen13} Kuang Shing Chen, Zi Yang Meng, Unjong Yu, Shuxiang Yang, Mark Jarrell, and Juana Moreno, Phys. Rev. B {\bf 88}, 041103(R) (2013).

\bibitem{Valkov15} V. V. Val'kov, T. A. Val'kova, and V. A. Mitskan, JETP Lett. {\bf 102}, 361 (2015).

\bibitem {Feng0306} Shiping Feng, Phys. Rev. B {\bf 68}, 184501 (2003); Shiping Feng, Tianxing Ma, and Huaiming Guo, Physica C {\bf 436}, 14 (2006).

\bibitem {Feng15} See, e.g., the review, Shiping Feng, Yu Lan, Huaisong Zhao, L\"ulin Kuang, Ling Qin, and Xixiao Ma, Int. J. Mod. Phys. B {\bf 29}, 1530009 (2015).

\bibitem{Zhao12} Huaisong Zhao, L\"ulin Kuang, and Shiping Feng, Physica C {\bf 478}, 49 (2012).

\bibitem{Feng0494} Shiping Feng, Jihong Qin, and Tianxing Ma, J. Phys.: Condens. Matter {\bf 16}, 343 (2004); Shiping Feng, Z. B. Su, and L. Yu, Phys. Rev. B {\bf 49}, 2368 (1994).

\bibitem{Feng15a} Shiping Feng, L\"ulin Kuang, and Huaisong Zhao, Physica C {\bf 517}, 5 (2015).

\bibitem{Zhou08} Sen Zhou and Ziqiang Wang, Phys. Rev. Lett. {\bf 100}, 217002 (2008).

\bibitem{Honerkamp03} Carsten Honerkamp, Phys. Rev. B {\bf 68}, 104510 (2003).

\bibitem{Watanabe04} Tsutomu Watanabe, Hisatoshi Yokoyama, Yukio Tanaka, Jun-ichiro Inoue, and Masao Ogata, J. Phys. Soc. Jpn. {\bf 73}, 3404 (2004).

\bibitem{Weber06} C\'edric Weber, Andreas L\"auchli, Fr\'ed\'eric Mila, and Thierry Giamarchi, Phys. Rev. B {\bf 73}, 014519 (2006).

\bibitem{Kiesel13} Maximilian L. Kiesel, Christian Platt, Werner Hanke, and Ronny Thomale, Phys. Rev. Lett. {\bf 111}, 097001 (2013).


\end{thebibliography}
\end{document}